# Synthesis and scintillation properties of some dense X-ray phosphors

[1]Dujardin C., [1]Garcia-Murillo A., [1]Pedrini C., [1]Madej C.,
[1]Goutaudier C., [2]Koch A., [3]Petrosyan A.G., [3]K.L.Ovanesyan, [3]G.O.Shirinyan, [4]Weber M.J.

[1]*Laboratoire de Physico-Chimie des Matériaux Luminescents, Unité Mixte de Recherche 5620 CNRS- Université Lyon1, 69622 Villeurbanne, France*
[2]*European Synchrotron Radiation Facility, BP 220, F-38043 Grenoble.*
[3]*Institute for Physical Research, Armenian National Academy of Science, 378410 Ashtarak-2, Armenia*
[4]*Lawrence Berkeley National Laboratory, Berkeley, CA 94720 USA*

**Abstract.** Many ultra-dense lutetium or gadolinium based compounds doped with $Eu^{3+}$ have been prepared. This paper reports on the major scintillation performances of these compounds. One of them ($Lu_2O_3$:Eu) is particularly promising and have been deposited on a screen. Performances of such a screen are presented.

**Introduction**

X-ray phosphors have been developed since W.C.Roentgen discovered x-rays. Many systems are actually known and used in different devices for x-ray imaging systems. Nevertheless, detection technologies are improving their performance, and better x-ray converters would benefit image quality. For example, most of the x-ray phosphors have been developed for film radiography which are mainly green-blue sensitive. Actually, CCD camera as well as MOS technology which have good efficiency in the red range allow to obtain fast results without chemical processing. Requirements for such scintillation polycrystalline powders are:

- high effective atomic number ($Z_{eff}$) to obtain good stopping power of incident radiation
- good scintillation light yield
- afterglow must be avoided
- emission wavelength should match the maximum efficiency of the photodetector.

The most commonly used material for x-ray imaging systems is $Gd_2O_2S$:$Tb^{3+}$ (GOS). Its density is 7.34g/cm$^3$ ($Z_{eff}$=61.1) and its light yield is around 78000 photons/Mev depending on the granularity and preparation conditions. The major emission line is peaking at 547 nm (well-known $(4f^8)^5D_4 \rightarrow (4f^8)^7F_5$ green emission of $Tb^{3+}$). Most of the optical scintillation properties of this material are described in [1]. The time-constant of this emission line is in the ms range with low afterglow (0.1% if the initial signal within the 20ms after the x-ray beam obturation). For imaging with high spatial resolution thin powder phosphor layers are required because the spatial resolution is proportional to the thickness of the phosphor layer. If such a screen is used at high x-ray energies, the absorption and hence the image quality may be significantly reduced as a result of its thickness. Phosphors of higher density and higher effective atomic number than GOS improve the detector performance, under certain conditions even if the light yield is smaller. The purpose of this work was to find a polycrystalline materials having a better stopping power than GOS, and having other properties as good as GOS.

For the above applications research on very dense host materials doped essentially with $Eu^{3+}$ (well-known very efficient luminescence) have been investigated. Synthesis, spectroscopic and scintillation properties of $Lu_2O_3$, $LuTaO_4$, $Lu_3TaO_7$, $GdTaO_4$, $Gd_3TaO_7$ and $LiLu(WO_4)_2$ are presented.

**Material preparation and experimental set-up**

Europium-doped tantalates ($Gd_3TaO_7$, $GdTaO_4$, $Lu_3TaO_7$, $LuTaO_4$), oxides ($Lu_2O_3$, $Gd_2O_3$) and $LiLu(WO_4)_2$ tungstate were prepared by the solid state reaction technique. The mechanical mixed 3N to 4N purity oxide (except lithium carbonate) components were pressed into discs and fired in air with grinding and pressing procedures repeated at each successively higher temperatures. Tantalate compounds were also prepared by melting the mixed component oxides in Mo crucible under a reducing atmosphere with subsequent annealing in air. The germanate was prepared in sealed ampoules. The resulted materials were colorless and single phase. The lattice unit cell parameters measured by x-ray diffraction were close to those reported in [2,3].

Light yield have been measured by comparison with the light obtain with GOS from Riedel de Haen (RGS-N-Green). The x-ray source was operated at 15 kV, with Cu anode (the x-rays were mainly of 8 keV energy). The beam was collimated to 1 mm x 1.5 mm and illuminated the powder at 45° angle. The x-rays were mainly an energy of 8 keV. The powder has been deposited on aluminium plate; the thickness of the layer is 500 µm. With such a configuration, all the x-rays are absorbed. The signal was detected with a Si-photodiode perpendicular to the powder screen. The spectral response of the detector is 0.3 A/W at 550 nm and 0.35 A/W at 650nm.

|  | Zeff | $\rho$ (g/cm$^3$) | $Z_{eff}^4.\rho$ (rel) | Structure | Space group | emission (nm) |
|---|---|---|---|---|---|---|
| $Gd_2O_2S$ | 61.1 | 7.34 | 1 | hexagonal |  | 547 |
| $Gd_2O_3$ | 61.8 | 7.62/ 8.33 | 1.09/ 1.22 | cubic/ monoclinic | Ia3 C2/m | 611.2 614.9/623.2 |
| $Gd_3TaO_7$ | 64.3 | 8.42 | 1.44 | rhombic | $D_2^5$ | 612/615 |
| $Lu_2GeO_5$ | 65 | 7.53 | 1.31 | monoclinic | $C_{2h}^6$ | 611 |
| $GdTaO_4$ | 66.3 | 8.83 | 1.7 | monoclinic | P2/a | 612.3/615.4 |
| $Lu_2O_3$ | 68.8 | 9.42 | 2.11 | cubic | Ia3 | 611.1 |
| $Lu_3TaO_7$ | 68.9 | 9.49 | 2.09 | cubic | $O_h^5$ | 612.3/615.4 |
| $LiLu(WO_4)_2$ | 69.1 | 7.92 | 1.81 | monoclinic | P2/n | 609/broad 480 |
| $LuTaO_4$ | 69.1 | 9.76 | 2.22 | Monoclinic | P2/a | 611.9 |

Table 1: Properties of the host materials and $Eu^{3+}$ doped materials

**Results**

Neglecting the absorption edge, the x-ray absorption efficiency $\eta_{abs}$ depends upon the effective number ($Z_{eff}$) and the density of the material such that: $\eta_{abs} \propto Z_{eff}^4.\rho/E_x^3$ [4] ($E_x$ is the x-ray energy). $Z_{eff}^4.\rho$ is reported in Table 1 relatively to GOS for comparison. Emission spectrum was recorded under 253nm excitation (charge transfer states of $Eu^{3+}$). As expected, red emission lines due to the $^5D_0 \rightarrow {}^7F_J$ transitions of $Eu^{3+}$ are observed. Major lines are listed in table 1. Spectroscopic behavior of $LuTaO_4:Eu^{3+}$ have been studied by Blasse et al. [5]. For $Lu_2O_3$ and $Gd_2O_3$ doped with $Eu^{3+}$ ref [6] and [7] give the main spectroscopic properties. The tungstate materials exhibit in addition to the $Eu^{3+}$ lines a very broad band peaking at 480nm which is attributed to the $(WO_4)^{2-}$. The undoped materials give rise to the broad band alone.

| Sample | LY |
| --- | --- |
| LiLu(WO4)2:5-10%Eu | 2 |
| Lu3TaO7:5%Eu | 4 |
| Lu2GeO5:5%Eu | 6 |
| Gd3TaO7:5%Eu | 7 |
| Lu2O3:25%Eu | 9 |
| LuTaO4:5%Eu | 10 |
| Gd2O3:0.1%Eu | 10 |
| Gd2O3:10%Eu | 10 |
| Lu2O3:0.05%Eu | 11 |
| Gd2O3:20%Eu | 13 |
| GdTaO4:1%Eu | 14 |
| GdTaO4:15%Eu | 16 |
| GdTaO4:3%Eu | 16 |
| Gd2O3:15%Eu | 18 |
| Gd2O3:0.5%Eu | 19 |
| Gd2O3:7%Eu | 19 |
| Gd2O3:3%Eu | 20 |
| Gd2O3:5%Eu | 20 |
| GdTaO4:10%Eu | 22 |
| GdTaO4:5%Eu | 22 |
| GdTaO4:1%Eu | 22 |
| GdTaO4:10%Eu | 23 |
| Lu2O3:0.25%Eu | 24 |
| Lu2O3:10%Eu | 28 |
| Lu2O3:0.5%Eu | 33 |
| Lu2O3:1.5%Eu | 38 |
| Lu2O3:2.5%Eu | 46 |
| Lu2O3:5%Eu | 47 |
| Gd2O2S:Tb (RGS-N) | 100 |

Table 2: list of the mono-doped samples and their relative light output

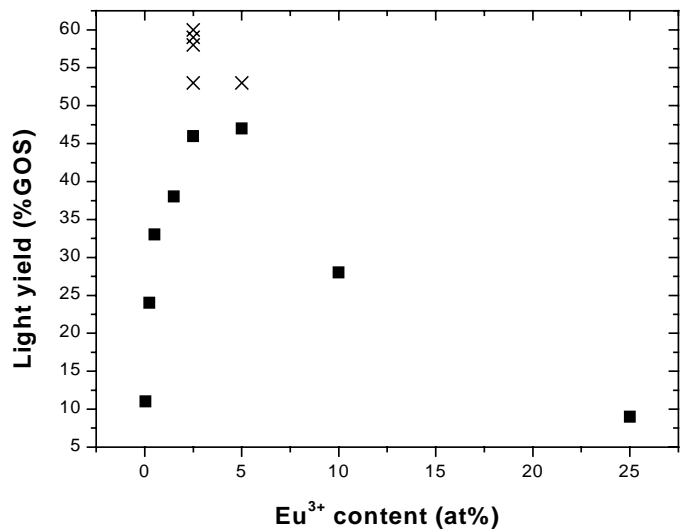

Figure 1: Light yield of $Lu_2O_3$ relative to GOS versus $Eu^{3+}$ concentration. Squares indicate the single-doped materials, and crosses are $Tb^{3+}$ co-doped compounds. The co-doping level is between 0.005 at. % and 0.1 at. %

The relative light yield was measured for many samples and listed in Table 2. The samples are sorted using the relative light yield value. For $GdTaO_4$, $Gd_2O_3$ (monoclinic) and $Lu_2O_3$, several concentrations have been tested. For $Lu_2O_3$, the variation of light yield versus the $Eu^{3+}$ content (atomic % ) is plotted in Figure 1. The two best single-doped samples are 2.5% and 5%. Extremely weak co-doping with $Tb^{3+}$ (between 0.005 at. % and 0.1 at. %) improves the light yield significantly as shown in Figure 1 (crosses). The origin of this effect is under study using VUV synchrotron radiation. Since divalent europium and tetravalent terbium are also stable energy transfer from host materials to both ions may develop in different ways and then $Tb^{3+}$ may capture remaining excitations and transfer its excitations to $Eu^{3+}$ or give rise to the green luminescence. However, no green luminescence have been detected (Figure 2).

### Screen test

$Lu_2O_3$:Eu,Tb already looks very promising although no optimization of the granularity or preparation conditions have been performed yet. Screens have been sedimented on 1 mm thick glass plates using this $Lu_2O_3$:Eu,Tb phosphor and the commercial $Gd_2O_2S$:Tb (RDS-N), 24 mg/cm$^2$ and 30 mg/cm$^2$ respectively. This deposition gave equal layer thickness of 50 µm and hence similar spatial resolution properties are expected. The screens have been characterized using an x-ray generator source with a Mo anode and 100-µm Zr filter operated at 30 kV, thus providing x-ray emission mainly at 18 keV. The spectral response is shown in Figure 2. The linespread function was measured to be 35 µm fwhm, in both cases. The non-uniformity is 14% rms for the $Lu_2O_3$ screen that is a factor of 2 higher than for the GOS screen when scanned with microscope optics of 1.2 µm pixel size at the screen. This non-uniformity is a result of the granularity. A larger pixel size would average and reduce these spatial variations.

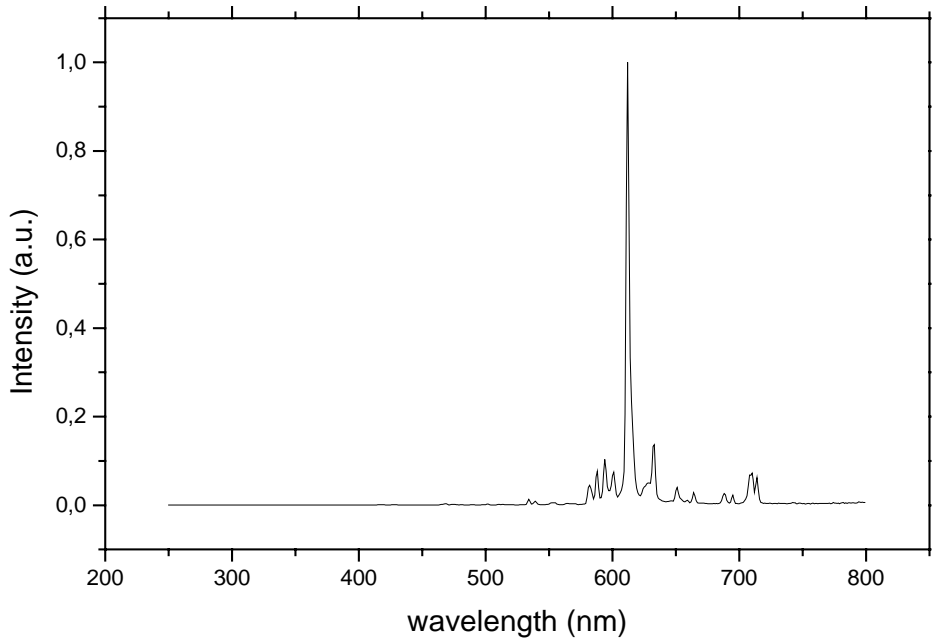

Figure 2: emission spectrum of $Lu_2O_3$:2.5%$Eu^{3+}$-0.005%$Tb^{3+}$ under x-ray excitation. High voltage is 15kV, Cu-anode, 25µmCu, 50µm Al

The afterglow has also been studied at exposure levels of $10^6$ ph/s/mm$^2$ (Fig. 3). The initial decay is slower than for GOS. After 1 s, a dynamic range of $10^3$ is possible. For even higher dynamic range, GOS is limited by afterglow; $Lu_2O_3$:Eu decays further. The decay curves additionally depend on the exposure time. A reduction of the afterglow may be achieved with different co-doping without reducing the light yield. Radiation damage has been observed at high doses: 1% reduction in the signal response after an absorbed dose of 30 Gy for $Lu_2O_3$:Eu and 100 Gy for $Gd_2O_2S$:Tb. The material recovers partially at room. The red emission is better adapted to the spectral response of front side illuminated CCD's than the green emission of GOS. Even though the conversion efficiency is lower than for GOS, a typical front side illuminated CCD does register similar signal amplitude for the same number of absorbed x-ray photons. The advantage compared to GOS is that the absorption of such a 50 µm thick screens is higher by 30% at 18 keV and 70% at 30 keV. In particular cases for which the x-ray shot noise dominates the detector performance, only the number of absorbed x-ray photons determines the signal-to-noise ratio and not the signal conversion efficiency of x-rays into an electronic signal.

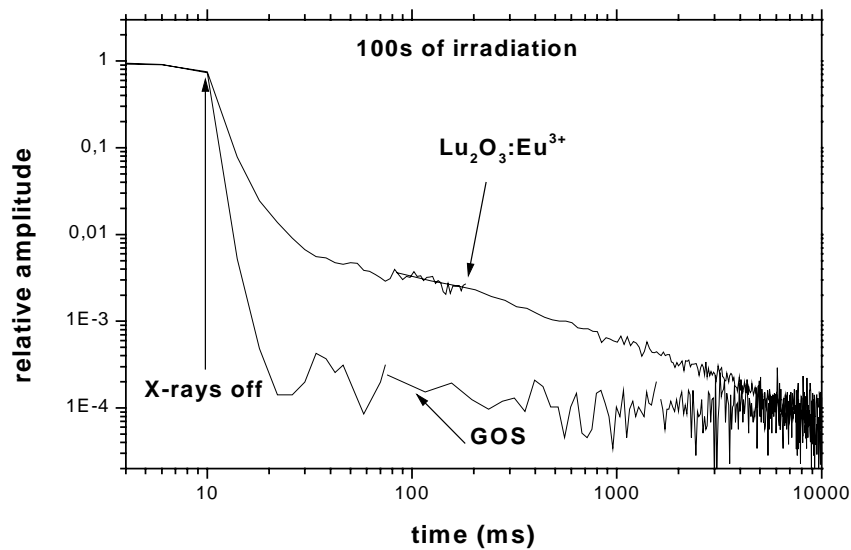

Figure 3 Afterglow measurement on GOS and $Lu_2O_3$:2.5%$Eu^{3+}$-0.005%$Tb^{3+}$ under X-ray excitation

**Conclusion**

Many very dense phosphors doped with $Eu^{3+}$ ions have been synthetised and studied. Of these, $Lu_2O_3$:Eu gives the best light yield. Additional co-doping with Tb improves significantly the light yield. The material and screen properties of $Lu_2O_3$:Eu are sufficiently advanced so that a higher signal-to-noise ratio in X-ray imaging under typical medical imaging conditions can be achieved compared to $Gd_2O_2S$:Tb. Hence, lower doses to patients are expected with $Lu_2O_3$:Eu screens. Comparative tests on detectors in a routine working environment are planned.